\documentclass[prl,aps,twocolumn,superscriptaddress,showpacs]{revtex4}

\usepackage{graphicx}

\begin{document}

\title{One-dimensional magnetic fluctuations in the spin-2 triangular lattice $\alpha$-NaMnO$_{2}$}

\author{C. Stock} 
\email{chris.stock@stfc.ac.uk; Correspondence}
\affiliation{ISIS Facility, Rutherford Appleton Laboratory, Didcot, Oxon, OX11 0QX, UK}
\author{L.C. Chapon}
\affiliation{ISIS Facility, Rutherford Appleton Laboratory, Didcot, Oxon, OX11 0QX, UK}
\author{O. Adamopoulos}
\affiliation{Institute of Electronic Structure and Laser, Foundation of Research and Technology-Hellas, Vassilika Vouton, 71110 Heraklion, Greece}
\affiliation{Department of Chemistry, University of Crete, Voutes, 71003 Heraklion, Greece}
\author{A. Lappas}
\email{lappas@iesl.forth.gr; Correspondence}
\affiliation{Institute of Electronic Structure and Laser, Foundation of Research and Technology-Hellas, Vassilika Vouton, 71110 Heraklion, Greece}
\author{M. Giot}
\affiliation{ISIS Facility, Rutherford Appleton Laboratory, Didcot, Oxon, OX11 0QX, UK}
\affiliation{Institute of Electronic Structure and Laser, Foundation of Research and Technology-Hellas, Vassilika Vouton, 71110 Heraklion, Greece}
\author{J.W. Taylor}
\affiliation{ISIS Facility, Rutherford Appleton Laboratory, Didcot, Oxon, OX11 0QX, UK}
\author{M.A. Green}
\affiliation{NIST Center for Neutron Research, 100 Bureau Drive, Gaithersburg, MD 20899-6102, USA}
\affiliation{Department of Materials Science and Engineering, University of Maryland, College Park, MD 20742-2115, USA}
\author{C.M. Brown}
\affiliation{NIST Center for Neutron Research, 100 Bureau Drive, Gaithersburg, MD 20899-6102, USA}
\author{P. G. Radaelli}
\affiliation{ISIS Facility, Rutherford Appleton Laboratory, Didcot, Oxon, OX11 0QX, UK}

\date{\today}

\begin{abstract}

The S=2 anisotropic triangular lattice $\alpha$-NaMnO$_{2}$ is studied by neutron inelastic scattering. Antiferromagnetic order occurs at T $\leq$ 45 K with opening of a spin gap. The spectral weight of the magnetic dynamics above the gap ($\Delta \simeq$ 7.5 meV) has been analysed by the single-mode approximation. Excellent agreement with the experiment is achieved when a dominant exchange interaction ($|J|/k_{B} \simeq$73 K), along the monoclinic $b$-axis and a sizeable easy-axis magnetic anisotropy ($|D|/k_{B} \simeq$3 K) are considered. Despite earlier suggestions for two-dimensional spin interactions, the dynamics illustrate strongly coupled antiferromagnetic S=2 chains and cancellation of the interchain exchange due to the lattice topology.  $\alpha$-NaMnO$_{2}$ therefore represents a model system where the geometric frustration is resolved through the lowering of the dimensionality of the spin interactions.

\end{abstract}


\maketitle

	In geometrically frustrated magnets, each spin cannot satisfy all pairwise interactions and therefore remains disordered to temperatures well below the Curie-Weiss temperature ($\Theta_{CW}$) where magnetic order is expected.~\cite{Collins97:75}  The ground state degeneracy imposed by lattice topology may lead to unconventional magnetic properties, including spin-liquid and nematic phases.~\cite{Anderson73:8,Tsun06:75}   Examples of two-dimensional (2D) triangular antiferromagnets include the Cs$_{2}$CuCl$_{4}$ (S=1/2), $\kappa$-(BEDT-TTF)$_{2}$Cu$_{2}$(CN)$_{3}$ (S=1/2), ZnCu$_{3}$(OH)$_{6}$Cl$_{2}$ (Kagome) and NiGa$_{2}$S$_{4}$ (S=1).~\cite{Coldea03:68,Nakatsuji05:309,Shimizu03:91,Helton07:98,Lee07:6}  All of these systems display anomalous magnetic properties well below $\Theta_{CW}$ and illustrate the dramatic effects that crystal symmetry imposed degeneracy, or geometric frustration, can have on the magnetic structures and excitations.

\begin{figure}[t]
\includegraphics[width=75mm]{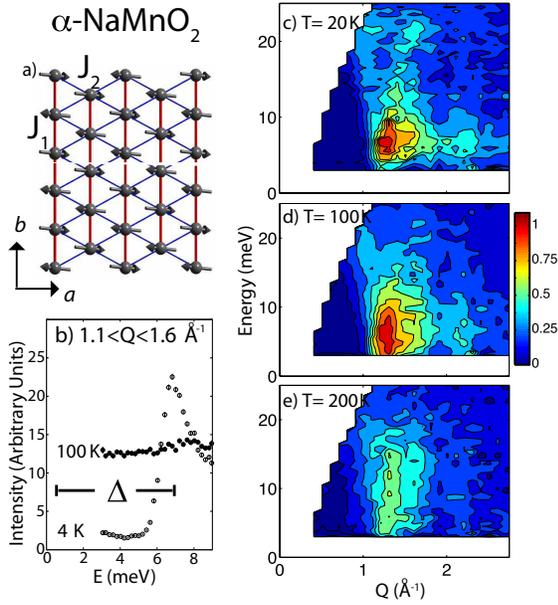}
\caption{ (a) Mn$^{3+}$ ions (projected in the basal plane); the thick red lines illustrate the chains, with $J_1$ (red) and $J_2$ (blue) marking the intra- and inter-chain interactions.  (b) Constant-Q cuts taken on the DCS spectrometer.(c-e) $T$-evolution of MARI spectra illustrate the strong magnetic fluctuations at higher temperatures.} \label{figure1}
\end{figure}

	$\alpha$-NaMnO$_{2}$ consists of manganese (Mn$^{3+}$, 3$d^4$; S=2) planes  that form an anisotropic triangular lattice.~\cite{Giot07:99}  The magnetic and crystal structures have been investigated by neutron powder diffraction (NPD).
This has revealed an interplay between a N\'{e}el state and strong 2D diffuse scattering, indicative of short-range spin correlations. The underlying structure of the Mn$^{3+}$ spins is illustrated in Fig. \ref{figure1}a. It is composed of edge-shared isosceles Mn-triangles, with short interatomic distances (2.86 \AA) along the monoclinic $b$ cell direction and longer ones (3.17 \AA) along the cell diagonal. This distortion results in nonequivalent Mn-Mn exchange pathways. The Jahn-Teller distorted Mn cations form antiferromagnetic (AFM) chains parallel to the $b$-axis, while the moment direction is fixed along the $d_{3z^{2}-r^{2}}$ orbital order (OO). The lattice topology could then support different intrachain ($J_1$) and interchain ($J_2$) exchange integrals.
	 
	The magnetic interactions are assumed highly 2D as evidenced through a combined susceptibility and ESR study, finding the ratio of interactions $J_{2}$/$J_{1}$ $\sim$ 0.44.~\cite{Zorko08:77} The 2D picture in $\alpha$-NaMnO$_{2}$ is in accord with expectations based on the isostructural NaNiO$_{2}$ (S=1/2), which displays strong ferromagnetic 2D interactions within the $ab$ plane.~\cite{Lewis05:72,Darie05:43} In addition, NPD reports that 2D spin correlations develop below 200 K, with long-range (3D) magnetic order occuring simultaneously with a triclinic to monoclinic structural distortion at $T_N\approx$45 K.~\cite{Giot07:99} In the 3D ordered phase the Mn-chains order AFM in the basal plane and ferromagnetically along the $c$-axis.  Although the degeneracy of the ground state can be lifted through the magneto-elastic phase transition, which renders all Mn-Mn bonds in the triangles inequivalent, $\alpha$-NaMnO$_{2}$ is far from classical as evidenced by the low N\'{e}el temperature (45 K) compared with a $\Theta_{CW}$=490 K. 

	We have investigated the spin interactions in $\alpha$-NaMnO$_{2}$ using neutron inelastic scattering (INS) and demonstrate that the low-temperature magnetic excitations are surprisingly one-dimensional (1D); the values of the intrachain exchange as well as that of the magnetic anisotropy are determined.  The former result is in contrast to expectations based on earlier studies outlined above,~\cite{Zorko08:77,Giot07:99} which all point to strong 2D interactions.  We argue that the combined effect of degeneracy introduced through geometric frustration and the OO result in the cancellation of the interchain interactions and effectively lower the dimensionality of $\alpha$-NaMnO$_{2}$.

	Experiments were conducted on MARI (ISIS) and the Disk Chopper Spectrometer (DCS, NIST).  On MARI, neutrons with incident energies ($E_i$) of 30, 85, and 150 meV were used at selected chopper frequencies, resulting in energy resolutions of 1.2, 4.2, and 5.8 meV,  respectively.  Data was corrected for a constant and phonon background by fitting the intensity in the high angle banks (where magnetic scattering is absent) to the form $I_{Back}=B_{0}+B_{1}Q^{2}$ and extrapolating to small momentum transfer.  The assumptions behind this correction break down for elastic energy transfers and therefore we have removed the elastic line from the plots. A V-standard of known mass was used to perform absolute calibration.  The intensity is related to the powder average scattering function, $S(Q,E)$, through $\tilde{I}(Q,E)=(\gamma r_{0})^2 |(g/2)F(Q)|^{2}2S(Q,E)$), where $(\gamma r_{0})^2=0.29$ barn sr$^{-1}\mu_B^{-2}$, $F(Q)$ is the magnetic form factor, and $g$ is the Land\'{e} factor. On DCS, $E_i$ was set to 12.1 meV with a full-width resolution of 0.73 meV. The single phase, air sensitive, powder (9.6 g at ISIS and 6.6 g at NIST mounted in annular Al can) was synthesised as reported~\cite{Parent71:3} and characterized at BT1 (NIST) with parameters (space group $C$2/$m$) $a$=5.670 \AA, $b$=2.855 \AA, $c$=5.804 \AA\ and $\beta$=113.23$^{\circ}$ at 300 K. 
	
	Figure \ref{figure1} provides an overview of the excitation spectrum and the temperature evolution.  Data taken on DCS (Fig. \ref{figure1}b) show the presence of a spin-gap of $\Delta\simeq$ 7.5 meV in the excitation spectrum at low-temperatures and the disappearance of this gap, yet the presence of significant magnetic scattering, at $T>T_N$.   The magnetic scattering measured with $E_{i}$=85 meV at 20, 100, and 200 K on MARI are illustrated in Fig. \ref{figure1}c-e.  The spectrum at low-temperatures ($T \leq T_N$) shows the presence of a gap that opens up abruptly and dispersing spin-waves at $E>\Delta$.  Strong magnetic fluctuations still persist well above $T_N$ (Fig. \ref{figure1}d-e), though the magnetic gap fills-in with temperature (Fig. \ref{figure1}b). 

\begin{figure}[t]
\includegraphics[width=80mm]{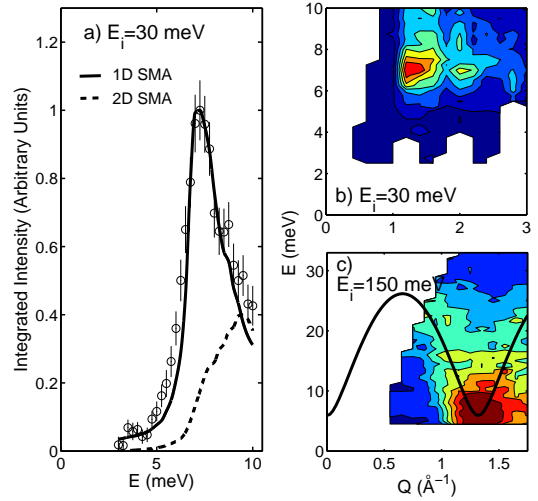}
\caption{ (a) The magnetic density of states integrated over the range of 1.0 $<$ $Q$ $<$ 2.5 \AA$^{-1}$ (T=5 K).  The solid and dashed lines are calculations based on the 1D and 2D SMA models, respectively. (b) The low-energy spectrum on MARI, with $E_{i}$=30 meV; the white spaces are due to gaps in the detectors.  (c) The excitation spectrum measured with $E_{i}$=150 meV illustrates the dispersion and the location of the top of the magnetic band.  The solid line is the single crystal dispersion, taking Q $\parallel$ to the Mn-chains. } \label{figure3}
\end{figure}

	Because the excitation spectrum (Figs. \ref{figure3}) appears to be dominated by a single coherent mode, we utilize the single-mode approximation (SMA) as a first step in understanding the spin dynamics.  The measured structure factor can be written in terms of a momentum dependent part and a Dirac delta function in energy $S(\vec{Q},E)=S(\vec{Q})\delta(E-\epsilon(\vec{Q}))$.  We have approximated $\delta(E)$ as a Lorentzian, with a full-width equal to the calculated resolution width.  The first moment sum~\cite{Hohenberg} relates $S(\vec{Q})$ to the dispersion $\epsilon (\vec{Q})$,

\begin{eqnarray}
S(\vec{Q})=-{2\over 3}{1\over \epsilon(\vec{Q})}\sum_{\vec{d}}J_{\vec{d}} \langle \vec{S_{0}} \cdot \vec{S_{\vec{d}}} \rangle [1-\cos(\vec{Q}\cdot\vec{d})].
\end{eqnarray}

\noindent Here $\vec{d}$ is the bond vector connecting nearest neighbour (NN) spins, with a superexchange interaction $J$.  This description of the dynamics is exact and only relies on the presence of only 1 Mn$^{3+}$ in the unit cell.~\cite{Hong06:74,Stone01:64,Xu96:54}  

	Figure \ref{figure3}a illustrates the momentum integrated magnetic intensity (Fig. \ref{figure3}b) weighted by the wavevector transfer $\tilde{I}(E)=\int dQ Q^{2} I(Q,E)/\int dQ Q^{2}$, which is a measure of the magnetic density of states and is sensitive to the dimensionality. The peak in the integrated weight is surprising as it implies that the correlations are not 2D, as expected from susceptibility and structural studies, ~\cite{Zorko08:77,Giot07:99} but rather strongly 1D.  To verify this, we use the SMA presented above, with a single exchange along the shortest Mn-Mn interatomic distance $b$, (ie $J$= $J_1$ in Fig. \ref{figure1}a).  The 1D SMA  calculation is compared with the low-energy spectrum in Fig. \ref{figure3}a. We also show a representative calculation assuming a weak 2D interaction taken to be 10\% of $J$.  While the 1D model provides an excellent description of the data, the weakly 2D model fails to reproduce the sharp peak in the $\tilde{I}(E)$.

\begin{figure}[t]
\includegraphics[width=85mm]{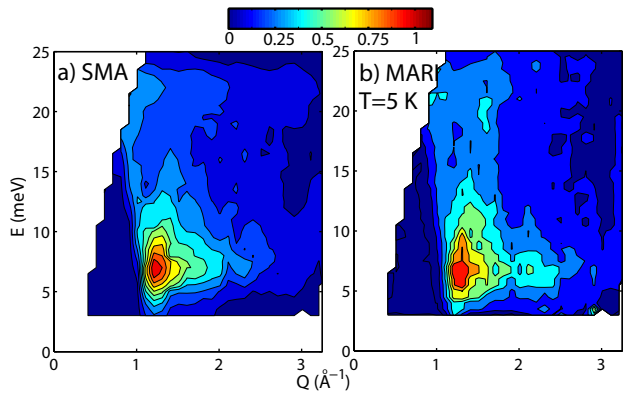}
\caption{(a) The calculated intensity contours for the 1D SMA described in the text.  (b) The measured magnetic intensity on MARI, with $E_{i}$=85 meV at T=5 K.} \label{figure2}
\end{figure}

To obtain a model dispersion, we conducted measurements with $E_{i}$=150 meV (T=5 K) to provide a better detector coverage of the lowest $\vec{Q}$ Brillouin zone (Fig. \ref{figure3}c).  Due to the powder average, in a constant-Q cut at low momentum transfers, the energy onset of the magnetic scattering is sensitive to the dispersion.  Following the postulate that 1D interactions are dominant we use the dispersion relation for a 1D chain system,~\cite{Hong06:74,Stone01:64}

\begin{eqnarray}
\epsilon^{2}(\vec{Q})=4S^{2}(D^2+2JD+J^2\sin^2(\vec{Q}\cdot\vec{d})),
\end{eqnarray}

\noindent where $D$ is the magnetic anisotropy and $J$ is the NN exchange coupling.  The powder averaged dispersion, which defines the onset of magnetic scattering was calculated, with $|D|/k_B$=3.0 $\pm$ 0.5 K and $|J|/k_B$=73 $\pm$ 5 K and depicted in Fig. \ref{figure3}c. Both $|D|$ and $|J|$ are in agreement with those derived from earlier ESR and SQUID experiments~\cite{Zorko08:77} and confirm that the gap at $T \leq T_N$ is due to a sizable single-ion anisotropy. Quantum effects, such as the Haldane conjecture where a gap ${\Delta}'\simeq 0.0876 J$ is predicted for S=2 chains,~\cite{Wang99:60} are less likely to dictate the low-$E$ spectrum. These become significant for lower integral spins, such as for the spin-1 1D antiferromganet CsNiCl$_{3}$.~\cite{Buyers86:56} In addition, our INS data point to an anisotropy arising from an easy-axis as opposed to an easy-plane. The latter would result in a gapless mode in the excitation spectrum~\cite{Tranquada89:40} which is not evident here.

Using these parameters we calculate the magnetic response over the entire band and compare it with the measured intensity (Fig. \ref{figure2}). The overall dispersion at low-Q, the gap, and the bandwidth of the excitation spectrum are all captured in this 1D model. 

\begin{figure}[t]
\includegraphics[width=85mm]{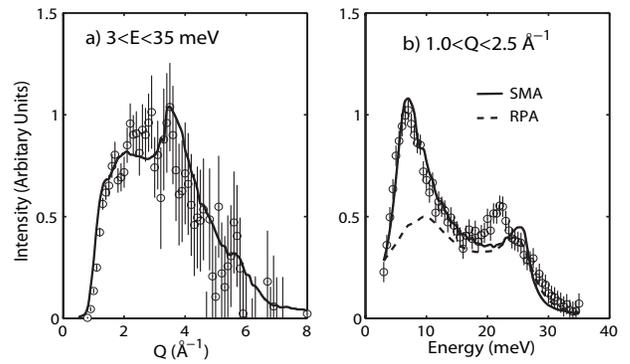}
\caption{The momentum (a) and the energy (b) integrated intensities, with E$_{i}$=85 meV at T=5 K.  Solid lines in both panels are the results of the SMA calculation. The dashed line in panel (b) is a calculation introducing inter-chain interactions through RPA.} \label{figure4}
\end{figure}

	The 1D magnetic properties of $\alpha$-NaMnO$_{2}$ and the valididty of the AFM chain model was further tested by analyzing the momentum and energy integrated intensities (T=5 K).  The latter ($I \propto \int dE Q^{2} I(Q,E)$) is illustrated in Fig. \ref{figure4}a and is sensitive to both the dimensionality and also the distance ($\vec{d}$) between correlated spins.  The onset observed at $Q_{0} \sim $1.2 \AA$^{-1}$ (Fig. \ref{figure2}a)  indicates that the dominant correlated spins are separated by a distance $d \sim \pi/Q_{0} =$ 2.6 \AA . For comparison, the spacing between Mn$^{3+}$ ions along the $b$ direction is 2.86 \AA, while the next NN Mn$^{3+}$ is at 3.17 \AA\ ($J_2$-pathway).  The steep rise in Fig. \ref{figure4}a  also contrasts the diffuse 2D Warren~\cite{Warren41:59} scattering observed at the elastic line and previously attributed to 2D correlations. The calculated intensity (Fig. \ref{figure4}a) from the 1D SMA, with $\epsilon (Q)$ of Eq. (2), is in accord with the data and thus supports the S=2 chain model.  

	The experimental Q-integrated intensity ($I \propto \int dQ Q^{2} I(Q,E)$) is described similarly well by the 1D SMA model (solid line in Fig. \ref{figure4}b).  It is important to investigate the effects of next NN interactions both within the chains and between chains.  While these affect the excitations near the top of the dispersion band (~\cite{Coldea01:86,Stock07:75,Kim01:64}), they also have significant effects at low-$E$ and can be ruled out.~\cite{Zheludev00:62}  We have introduced interactions between chains using the Random Phase Approximation (RPA).~\cite{Zheludev00:62} The dashed line in Fig. \ref{figure4}b is a RPA calculation taking $J_{2}/J_{1} \sim 0.1$; it does not describe the data.  Next NN exchange within a chain has significant effect on $\Delta$.  Given we find good agreement with other techniques for the anisotropy, we conclude that these interactions are negligible. Thus, excitations in $\alpha$-NaMnO$_{2}$ are dominated by a single exchange interaction along $b$.  

	While SMA describes the sharp peak at low energy transfers, deviations appear at higher energies near the top of the magnetic excitation spectrum (at $\sim$ 20 meV).   This peak is unlikely associated with the underlying physics, but rather to experimental error in subtracting the background from the small scattering angle banks and in particular near $E\sim$ 20 meV, where strong phonons in the sample and Al can exist.~\cite{Kresch08:77} 

	We now investigate the total integrated intensity which is governed by the zeroeth moment sum-rule, $I=\int dE \int d^{3}Q S(Q,E)/\int d^{3}Q=S(S+1)$.~\cite{Hammar98:57}  The integral is over all energies including the elastic magnetic Bragg peak which contributes $g^{2}\langle S_{z} \rangle^{2}$ (with g=2 for Mn$^{3+}$).  NPD has found an ordered moment of $\sim$ 3 $\mu_{B}$ implying $\langle S_{z} \rangle^{2}$ $\sim$ 2.3,~\cite{Giot07:99} while integration of INS gives 3.2 $\pm$ 0.4.  The total moment sum is therefore $I$ $\sim$ 5.5 and within error of the expected value of 6. This indicates we have captured the bulk of the low-$T$ spectral weight and there is little magnetic scattering residing in any diffuse, quasielastic or high-energy component.   It is also suggestive that the missing spectral weight inferred from NPD~\cite{Giot07:99} resides in the INS channel.

	The 1D excitations in $\alpha$-NaMnO$_{2}$ are surprising given the 2D behavior of Refs. \onlinecite{Giot07:99,Zorko08:77} and INS studies in isostructural compounds~\cite{Lewis05:72}.  The former indirect methods of  characterizing the dynamics were conducted above $T_{N}$ and probe very different time domains than studied here.  Future work investigating the $T$-dependence and low-energy response will be important for understanding these differences.
 
	While spin excitations along the monoclinic $b$ axis (Fig. \ref{figure1}a) are dispersive and cost energy due to the strong direct exchange, excitations perpendicular to the chains are degenerate and cost no energy given the triangular topology of the spins as imposed by the crystal symmetry. Therefore, $\alpha$-NaMnO$_{2}$ is a representative model system where the degeneracy introduced by the anisotropic lattice results in an effective lowering of the dimensionality of the magnetic excitations.  This mechanism is similar to the symmetry cancellation proposed to occur in LiMnO$_{2}$ at high-$T$ and theoretically suggested to explain spin lattice coupling in pyrochlores.~\cite{Greedan97:128,Tcher02:88}  

	In other triangular lattices, such as NiGa$_{2}$S$_{4}$ and Cs$_{2}$CuCl$_{4}$, geometric frustration is resolved through incommensurate or spiral order of the spins.  Such phases are precluded in $\alpha$-NaMnO$_{2}$ as OO fixes the moment direction and stabilizes a collinear state.  The combination of the triangular geometry and the OO in $\alpha$-NaMnO$_{2}$ results in excitations perpendicular to the chain direction to be degenerate in energy and necessitate the ground state to be 1D. 

	We have mapped out the magnetic excitation spectrum of the anisotropic $\alpha$-NaMnO$_{2}$ triangular antiferromagnet.  The data contradict expectations for strong 2D coupling and are described by a 1D antiferromagnetic chain model dominated by a single mode.  The derived parameters of the Hamiltonian, include the intrachain exchange ($|J|/k{_B}\simeq$73 K) and the magnetic anisotropy ($\mid$D$\mid$/$k{_B}\simeq$ 3 K). Below the N\'{e}el ordering a spin-gap ($\Delta/k_{B}\simeq$87 K), due to easy-axis anisotropy, separates the ground state from the excitations.  We propose that the subtle balance of degeneracy introduced through lattice topology result in the effective lowering of the dimensionality of the spin excitations and cancellation the interchain interactions.
	
	We thank R. Coldea, R.Cowley and S.-H. Lee for discussions.  AL acknowledges financial support from the European Commission (``Construction of New Infrastructures", contract no. 011723).  We are thankful to NSF (No. DMR-0454672) for partial support at NIST.


\end{document}